\documentclass[letterpaper,twocolumn,10pt]{article}

\usepackage[T1]{fontenc}
\usepackage[pdftex]{graphicx}
\usepackage{amsmath}
\usepackage{ctable}
\usepackage[hyphens]{url}
\definecolor{darkgreen}{rgb}{0.00,0.39,0.00}
\usepackage[
    colorlinks,
    linkcolor=blue,
    citecolor=darkgreen,
    urlcolor=blue,
]{hyperref}
\usepackage{verbatim}
\usepackage{color}
\usepackage{xspace}
\usepackage{txfonts}
\usepackage{usenix}
\usepackage{epsfig}
\usepackage{endnotes}
\usepackage{cite}
\usepackage{paralist}
\usepackage{tablefootnote}
\usepackage{subfig}
\usepackage{multirow}
\usepackage{microtype}
\usepackage[font={small}]{caption}
\usepackage[binary-units]{siunitx}
\usepackage[normalem]{ulem}
\usepackage{hyperref}
\usepackage[compact]{titlesec}

\usepackage[ruled,vlined]{algorithm2e}
\SetKwInOut{Input}{Input}
\SetKwInOut{Output}{Output}
\SetKwBlock{Block}{}{}
\SetAlFnt{\small}

\usepackage[inline]{enumitem}
\usepackage{enumerate}
\setlist{nosep}

\usepackage{listings}
\definecolor{red3}{rgb}{0.80,0.00,0.00}

\newcommand{\pr}[1]{{\small\texttt{#1}}}

\newcommand{\CM}[1]{}

\def\sp{\hspace{0.2in}}

\begin{document}
\title{%
    Safe and Efficient Remote Application Code Execution on Disaggregated NVM Storage with eBPF
}

\author{%
\rm Kornilios Kourtis\footnotemark[2] \thanks{Majority of work done while at IBM Research} \sp
\rm Animesh Trivedi\footnotemark[3]\sp
\rm Nikolas Ioannou\footnotemark[4]\sp
\medskip\\
\footnotemark[2] Independent researcher\sp
\footnotemark[3] Vrije Universiteit (VU), Amsterdam \sp
\footnotemark[4] IBM Research, Zurich \sp
%
}

\date{}
\maketitle

\begin{abstract}
With rapid improvements in NVM storage devices, the performance bottleneck 
is gradually shifting to the network, thus giving rise to the notion of ``data
movement wall''. To reduce the amount of data movement over the network, 
researchers have proposed near-data computing by shipping operations and 
compute-extensions closer to storage devices. However, running arbitrary, 
user-provided extensions in a shared, disaggregated storage environment 
presents multiple challenges regarding safety, isolation, and performance. 
Instead of approaching this problem from scratch, in this work we make a case for 
leveraging the Linux kernel eBPF framework to program disaggregated NVM storage devices. 
eBPF offers a safe, verifiable, and high-performance way of executing untrusted, 
user-defined code in a shared runtime. In this paper, we describe our experiences 
building a first prototype that supports remote operations on storage using 
eBPF, discuss the limitations of our approach, and directions for addressing them.
\end{abstract}

\section{Introduction}

NVM disaggregation over network (e.g., NVMoF) 
offers a number of advantages
including better utilization and independent scaling of compute and
storage\cite{klimovic16:flash-disaggregation}.
To meet the performance 
and capacity demands, the underlying storage has innovated 
rapidly in the mediums, interfaces, and packaging (multiple
parallel banks and channels) to push the performance of NVM devices to the 
near-memory performance in a short span of time~\cite{citeulike:2739855,optane-eval,zssd}. 
In contrast, the connecting network (or, in general, the link technologies, e.g. PCIe, Ethernet, 
Fiber Channel) performance is improving at a slower pace. For example, 
a modern storage device can read upwards of 10-15 GB/sec~\cite{segate} which 
exceeds the bandwidth of a 100 Gbps network link (or a PCIe Gen 3 x8 
link\footnote{PCIe and Ethernet standardizations are multi-year projects 
before they become available everywhere. In contrast, storage bandwidths have 
been evolving at a much faster pace.}). This performance gap between a storage 
device and its access network creates a \textit{``data movement wall''}, 
which is shifting the performance bottleneck from storage to network\cite{istvan:vldb2017,splinter}.

Naturally, there are many optimization opportunities to reduce the pressure on
the network by pushing operations closer to the storage\cite{koo:micro17}.  For
example, read-modify-write operations (e.g., incrementing a counter in a
key-value store) can be implemented with a single (instead of two) network
request(s), while database queries can be partially offloaded to the storage to
save bandwidth and
energy\cite{do:sigmod13,Kim:2016:IPD:2839532.2840070,Jo:yoursql:2016}. These
opportunities are widely acknowledged, leading to attempts to exploit them
either for local\cite{willow:osdi14,Cho:2013:ADM:2464996.2465003} or
remote\cite{istvan:vldb2017,splinter} storage. Shipping compute to
storage is not novel in itself and has a long
history~\cite{1998-sigmod-idisk,1998-asplos-active-disks,2001-cmu-active-disks},
but existing approaches focus on specific operations and applications, and
in most cases operate above the block layer. A disaggregated block service,
however, is meant to serve different applications.
Each application has its own requirements, 
storage layouts, and implementation nuances that make it difficult to come up with specific 
operations that can serve all applications equally well.

\begin{figure*}[t!]
    \center
    \subfloat[Expected performance when performing a remote numerical increment operation Vs offloading.]{%
        \includegraphics[width=.45\linewidth]{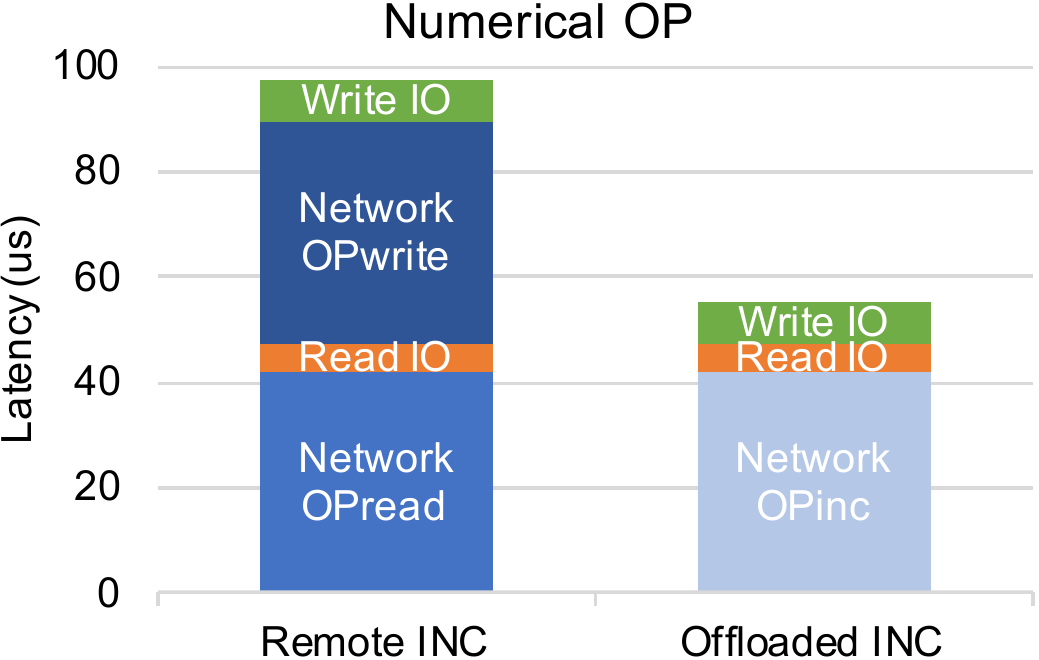}
        \label{fig:motiv-inc}
    }
    \hfill
    \subfloat[Expected performance of a remote binary search Vs offloaded binary search.] {%
        \includegraphics[width=.45\linewidth]{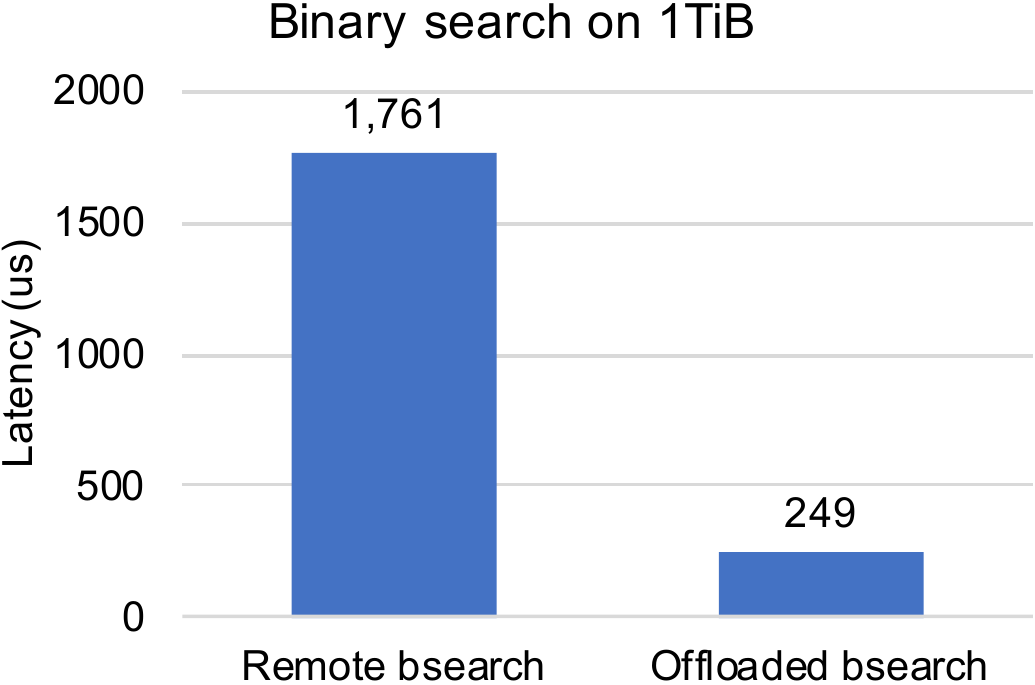}
        \label{fig:motiv-bsearch}
    }
    \caption{Examples of operations over remote data and their expected performance: (a) numerical increment and (b) binary search}
    \label{fig:motiv}
\end{figure*}

From a performance perspective, the ideal solution would be for the application to
upload native code to be executed on a storage service.  
However, programming storage devices specially in a shared and 
networked environment (i.e., disaggregated storage) is a challenging task 
for multiple reasons. First, a shared storage system needs to strike the right 
balance between expressiveness and safety
properties of extensions. 
Second, as performance of storage devices continues to improve rapidly, 
any extension on storage accesses must be executed with the utmost efficiency 
and minimum overheads. Lastly, from the usability point of view these extensions 
should be powerful and expressive and not tied to a particular application or 
domain.

In this paper, we explore using eBPF to solve  
the aforementioned challenges. The original BPF (BSD Packet Filter)\cite{bpf} 
was developed for implementing user-specific packet filters that could be safely 
and efficiently executed in the kernel, avoiding the need to copy all packets from 
kernel to user-space\cite{mccanne:bpf93}. Based on this, Linux has developed 
its own variant called eBPF\cite{ebpf,ebpf:starovoitov,cilium-bpf} that is used 
for a plethora of purposes.\footnote{\textit{``eBPF should stand 
for something meaningful, like Virtual Kernel Instruction Set (VKIS), but due to 
its origins it is extended Berkeley Packet Filter.''} - Brendan Gregg ~\cite{bg-ebpf}.}


\noindent\textbf{Why eBPF:}
eBPF is a good fit for three key reasons. 
First, unlike being tied to a single domain, the eBPF VM 
has emerged as an \textit{expressive} and the de facto way for user-space to 
push generic code into the Linux kernel, and is used in numerous applications: 
executing user-defined code in the network data-path before a packet enters 
the kernel network stack\cite{xdp,borkmann-xdp}, traffic control\cite{tc-bpf}, 
tracing\cite{bpf-tracing}, enhancing Linux security\cite{bpf-seccomp}, 
and writing safe kernel/hypervisor extensions~\cite{hyperupcalls}.
Second, to ensure \textit{safe} execution, the eBPF framework includes a
verifier to check correct and bounded execution of extensions. \textit{Performance} 
is delivered by using a JIT facility for compiling eBPF code into native 
(x86\_64, ARM) code. Lastly, eBPF is available now, and
is relatively mature and battle tested. Its wide adoption has resulted in 
a rich ecosystem of compilers, tools, and implementations that can be reused 
in other contexts~\cite{ebpf-1,ebpf-2,ebpf-dpdk,ebpf-bcc}, in addition to
the Linux kernel implementation.

Our approach is in contrast to a clean slate approach of implementing such a
facility from scratch (using proof-carrying code\cite{necula1996safe}, software
fault isolation (SFI)~\cite{sfi}, or safe language extensions like
Rust~\cite{splinter}). Indeed, we believe that if we had undertaken the latter,
we would have reached a fundamentally different design.  Starting from eBPF,
however, is a more pragmatic approach because it allows us to reuse a lot of
the existing infrastructure and benefit from the community built around it.

\noindent\textbf{Potential Gains:}
To motivate our work, we estimate the I/O operation overhead for doing two key operations over remote data:  (i) increment a numerical value and (ii) performing a binary search over a 1TiB sorted array of 8B values.
We measure the TCP round trip time of a small message (64B) over a 10GbE network using netperf (41.9$\mu$s), and use that to account for network I/O operations.
We measure the read and write latency (5.6 and 8.0$\mu$s, respectively) of an 3DXP NVMe drive, and use that to account for the read and write latency of storage I/O operations.
Fig~\ref{fig:motiv} illustrates the expected performance of these two operations when only accounting for I/O operations, not taking into account any data processing or contention.
Offloading the remote numerical increment operation can reduce the latency by 43\% (Fig~\ref{fig:motiv-inc}), due to halving the network operations required.
Offloading a binary search over 1 TiB of sorted 8B values ($2^{37}$ values) to the disaggregated storage provides even higher benefits, reducing latency by 86\% compared to performing this operation remotely (Fig~\ref{fig:motiv-bsearch}).
This large expected speedup can be attributed to the reduction of network operations from $37$ down to $1$.

In summary, reducing the network round trips helps to reduce the pressure on the network, thus 
improving an application performance. In this paper, we tackle this performance improvement problem
by using eBPF to ship to and safely execute code (which we call \emph{appcode}) on shared, 
disaggregated NVM devices.  

\section{eBPF Primer}



\noindent\textbf{Toolchain:} 
While it is possible to write eBPF directly in bytecode, the \pr{clang}
compiler includes a backend for generating eBPF from C code.  Each eBPF program
is defined as a function that takes a context (\pr{ctx}) argument and returns
an integer value. The type of \pr{ctx} depends on the type of the eBPF program
(e.g., packet filtering or profiling), as does the interpretation of the return
value (e.g., drop a packet when performing packet filtering).
Listing~\ref{lst:bpf} shows an example of such a program that drops all packets
that do not use IPv4.

\begin{lstlisting}[float,caption={C eBPF code example},label=lst:bpf]
int ipv4(struct xdp_md *ctx) {
  void *data_end = (void *)(long)ctx->data_end;
  void *data = (void *)(long)ctx->data;
  struct ethhdr *eth = data;
  if (data + sizeof(*eth) > data_end)
    return XDP_DROP;
  if (eth->h_proto == htons(ETH_P_IP))
    return XDP_PASS;
  return XDP_DROP;
}
\end{lstlisting}

\noindent\textbf{Verification:}
Because the eBPF code is provided by an untrusted application,
the kernel needs to ensure that the code execution is 
safe in two ways. First, it must not access (read or write)
arbitrary data. 
This is achieved by doing a form of symbolic execution where the verifier
goes over all possible execution paths and maintains information
about the range of values for the different registers.
Specifically, the verifier knows the size of the \pr{ctx} struct and can
determine whether an access based on \pr{ctx} is valid or not. Considering
Listing~\ref{lst:bpf}, the verifier can determine whether the accesses in lines
\pr{2} and \pr{3} fall within valid boundaries of the \pr{ctx} structure.
eBPF code may also access data outside of the \pr{ctx} structure
(e.g., raw network packet data), but the verifier needs to ensure that the
accesses are valid. This works by having fields of \pr{ctx} designating
valid regions. In the example of Listing~\ref{lst:bpf}, the region is within
\pr{ctx->data} and \pr{ctx-data\_end}.\footnote{There is also another region
the code is allowed to access: [{\ttfamily ctx->data\_meta}, {\ttfamily
ctx->data}], but we ignore this in the rest of the paper for the sake of
brevity.} The check in line \pr{5} ensures that the Ethernet packet header is
within the area that the eBPF code can access. Without that check, the verifier
would fail to validate the code on line \pr{7} since it cannot tell that this
is indeed a valid access.

Second, the execution time of eBPF code is bounded so that
other applications will not be denied service.
%
The verifier ensures that the eBPF function will return
in a reasonable amount of time by ensuring that there are no back-edges in the
control flow graph (disallowing loops), and that both the number of total
instructions but also the maximum number of instructions for any given control
flow path are bounded.



eBPF code is also allowed to invoke a number of helper
functions\cite{ebpf-helpers}. These functions are provided by the system as a
way to implement functionality that is not possible to implement within eBPF
itself, such as interacting with other subsystems.

\section{eBPF-powered Disaggregated NVM \mbox{Storage} Service Design}

\begin{lstlisting}[float,caption={Code sketch of task serving a client},label=lst:serve]
while (true) {
  recv(req); // receive request
  ty = req->type // request type
  if (ty == READ) {
    // read blocks from device
    rep = do_read(req);
  } else if (ty == WRITE) {
    // write blocks to the device
    rep = do_write(req);
  } else if (ty >= AC_BASE && ty < AC_MAX) {
    appcode = appcode_table[ty - AC_BASE];
    rep = appcode(req);
  } else rep = error();
  send(rep); // send reply
}
\end{lstlisting}

A disaggregated storage service provides different clients with storage devices
over the network via a protocol such as NVMoF\cite{nvmeof-spec},
AoE\cite{coile2005ata}, iSCSI, or NBD\cite{nbd-proto}\@.  Clients connect to a
given device, and after they are authenticated and authorized, they can issue
block-level commands (e.g., \pr{READ} or \pr{WRITE}) to it.
We built a prototype to explore using eBPF for a disaggregated NVM storage
service, which we briefly describe next. While our ideas are not necessarily
tied to this design, we include it because it allows for a more concrete
discussion.
The service serves requests over the network from multiple clients by issuing
appropriate commands to storage devices. 
%
%
%
Listing~\ref{lst:serve} (ignoring lines \pr{10}--\pr{12}) shows a
code sketch for a task that serves a client, issuing appropriate (e.g., read or
write) commands to the storage.  When an IO operation (e.g., \pr{recv},
\pr{read}, \pr{write}, or \pr{send}) is issued, the task switches to the
scheduler.  When the results of the requested IO operation become available, the
task will become runnable, and eventually be scheduled.  
To avoid the performance limitations
of synchronous IO\cite{c10k,barroso17}, we perform asynchronous IO using
collaboratively scheduled user-space tasks, as in our previous
work\cite{udepot:fast19}.
The tasks are written
so that they do not hold the CPU for a long time,
allowing other tasks to run and other clients to be served.  Moreover, using
appropriate algorithms in the scheduler, QoS policies can be implemented, but
this is beyond the scope of this work.

\subsection{Supporting eBPF appcode}

\begin{lstlisting}[float,caption={NBD request and reply headers},label=lst:nbd]
struct nbd_request {
    uint32_t magic;
    uint32_t type;   // == READ || == WRITE
    char handle[8];
    uint64_t from;  // offset (bytes)
    uint32_t len;   // length
};
struct nbd_reply {
    uint32_t magic;
    uint32_t error;  // 0 = ok, else error
    char handle[8];  // request handle
};
\end{lstlisting}

In our design, the eBPF verification and execution runs in the storage service
instead of the kernel. To this end, we ported the Linux kernel eBPF
infrastructure to user-space, and used it to build our prototype.

We use NBD in our prototype because it is simple (request and reply headers are
shown in Listing~\ref{lst:nbd}).  To support the eBPF appcode, we piggyback on NBD
by overloading the request type, and allow users to register different appcodes
and then invoke them by setting the proper request type
(Listing~\ref{lst:serve}, lines \pr{10}--\pr{12})

The appcode's argument (\pr{ctx}) in our prototype includes: a handle to the
storage device (e.g., file descriptor); data from the request header (\pr{from},
\pr{len}, and \pr{type}); and data pointers (\pr{data} and \pr{data\_end}) to
the payload of the request. The \pr{len} field contains the size of the payload,
while the \pr{from} field, which normally contains the operation offset in the
storage device, can be used arbitrarily by the appcode. The payload data are
also arbitrarily interpreted by the appcode.
The return value of the appcode is placed in the \pr{error} field of the reply,
while the appcode can also specify an area in the data region to be sent as
payload to the reply.

The appcode may use the following eBPF functions:
\pr{data\_realloc(s)},
which resizes the data area between \pr{data} and \pr{data\_end} to be
\pr{s} bytes, potentially allocating a new area and copying data;
\pr{io\_read(dev\_off,data\_off,s)}, which reads \pr{s} bytes from an
offset in the device (\pr{dev\_off}) and places them in \pr{ctx->data +
data\_off};
and \pr{io\_write(dev\_off,data\_off,s)}, that operates similarly to the read
operation, but writes data.

Using the above approach, it is possible to support remote operations on
disaggregated storage. Next, we describe two example use-cases: implementing
read-modify operations for a KV store, and SQL filter offloading.

\subsection{Example: remote increment operation}

Many data stores (e.g., key-value stores) support read-modify (e.g., increment)
operations\cite{memcached,redis}.  Implementing a read-modify operation in a
traditional disaggregated storage service requires two network requests. Using
our appcode facility we can perform the read-modify operation in the storage
server, requiring a single network request, reducing the latency and network
traffic.

We consider a key-value store as a concrete use case, with an in-memory hash table index.
Each hash table entry includes a key fingerprint,
the storage location of the full key-value record, and the size of the record.
The format of the key-value record in storage is as follows: \pr{(key length,
value length, key, value)}.
Increasing an integer value (with the same endianness as the host) requires:
\begin{enumerate*}[label=\roman*)]
\item fetching the key-value record from storage,
\item comparing the full key of the request with the key of the record,
\item increasing the value by one, assuming the keys match, and
\item writing the record back to storage.
\end{enumerate*}

We implemented this operation in our prototype.  The key-value store runs on
the client of the storage service, and invokes the appcode that performs the
increment. The \pr{from} field of the request includes the storage address of
the key-value record, while the payload includes the record size and the full
key. The appcode uses \pr{data\_relloc()} and \pr{io\_read()} to make space and
read the key-value record using the storage address and the record size. Then
it compares the key of the record with the key in the payload, and if they do
not match, it returns an appropriate error.  Otherwise, it increments the value
and uses \pr{io\_write()} to store it.  Because eBPF does not support loops,
the memory compare operations needs to be unrolled, so our approach can only be
applied to keys smaller than a certain size. If the key is larger than this
size, the key-value store works as it would have before.

%


%

\subsection{Example: Spark SQL filter offloads}
\label{sec:sql}

Here, we sketch how Spark SQL filters can be offloaded into an
eBPF-enhanced storage service for a ``near-device'' evaluation.
During the query planning phase, Spark generates row filters for selection
queries. A filter consists of a group of constraints on the value of columns in
a row. For example, in \pr{SELECT * FROM Customers WHERE CustomerID=1}, the
equality condition for the \pr{CustomerID} column is the filter. These filters
are then provided to an input data source, where the data are typically stored
using Parquet~\cite{pq} or ORC~\cite{orc}.  We assume a
disaggregated block device using one of these data formats to store data.

Many formats maintain metadata segments that contain column information.  For
example, in Parquet metadata are stored at the end of every data file.
Albis~\cite{albis}, another high-performance format, maintains metadata in a
separate file. The metadata contain information about the data such as minimum
and maximum values, sorted or not, all nulls or not, etc., that is used to
evaluate a filter, as well as the location of the data blocks.
These metadata are maintained in blocks of data ranging from 10 to 100MBs.
Instead of fetching these blocks to the client, we want to offload this
operation to the storage service by shipping appropriate eBPF code. The code
will receive parameters that specify the filters and the location of the
metadata block. Using this information, the code will read the metadata block,
and return the data blocks that the query needs to process (e.g., by considering
the minimal and maximum values of a column).

\subsection{Limitations and future directions}

Trying to implement the above examples (and also a Boyer-Moore string search
algorithm\cite{charras2004handbook}) in our prototype, we concluded that even
if these operations can be supported in a constrained form with our current
approach, the absence of loops in eBPF makes it difficult to support use cases
that involve processing large amounts of data.  Moreover, we found that the
compiler (\pr{clang}) has a hard time generating code that would pass the
verification process with high loop unroll counts.

The above issues are not surprising, since eBPF was not built with data
processing in mind. Nevertheless, we believe that this is an important use case
which should be supported.  In the next paragraphs, we discuss a number of
approaches for extending eBPF to address these issues.

\paragraph{Extend eBPF to support loops}
As eBPF is increasingly used, the inability to verify code with loops has
become a noticeable problem. Recent work, for example, aims to allow bounded
loops in eBPF\cite{ebpf-loops}. Although such a facility is certainly useful,
applications that operate on large amounts of data will still be limited in
what they can do.
A different approach would be to "teach" the verifier that some eBPF functions
(e.g, \pr{io\_read}, \pr{io\_write}, in our case) return the control to the
system, so they effectively reset the time the appcode is executing because the
system (the task scheduler in our case) may defer execution to other tasks.
This, however, requires substantial changes to the verifier, because
introducing loops has implications not only for bounding execution time, but
also for proving the safety of the code since now register values might be set
from multiple dataflow paths. Even though projects like KLEE\cite{klee} push
the boundaries of what symbolic execution can do, this is still a challenging
problem.

\paragraph{Total programs}

A program is total if it always terminates. It is a well-known theoretical
result that it is impossible to prove totality for Turing-complete languages,
but there are languages which sacrifice their generality to be able to prove
that their programs (assuming they compile) will always terminate.  One
approach would be to try and incorporate methods from this domain into eBPF to
allow for a richer set of computations for eBPF code.  This, however, is not
sufficient for practical purposes since a program might be guaranteed to
terminate, but still take a very long time to terminate. Indeed, the Dhall
language which is a configuration language following this approach, identifies
this issue in its documentation\cite{dhall-turing}.


\paragraph{High-order functions}

Another approach is to enrich the execution environment with high-order
functions such as \pr{map}, \pr{reduce} which may take user-defined functions
implemented in eBPF as arguments. This will result in a declarative approach of
the remote code to be executed, which allows the run-time system to decide how to
exactly execute it. At the same time, the user-defined functions, written in
eBPF, are verified for safety in terms of data access and bounded execution
times. The challenge with this approach is to determine the proper functions
that allow expressing the required programs, while at the same time allowing for
an efficient execution by the run-time system. We plan to follow this direction
in our future work.


%
%
%
%

\section{Related Work}

Pushing computation to storage is not a new idea and has a long history in systems 
like iDisk~\cite{1998-sigmod-idisk}, Active disks from UCSB~\cite{1998-asplos-active-disks}, 
and CMU~\cite{2001-cmu-active-disks}. 
Later systems like MapReduce~\cite{2004-odsi-mapreduce} 
and Spark~\cite{Zaharia:2012:RDD} popularized the idea of ``shipping code to data'' where compute 
was shipped to co-located data and compute nodes. Systems like YourSQL~\cite{Jo:yoursql:2016}, 
Ibex~\cite{Woods:ibex:2014}, and other works~\cite{do:sigmod13,Kim:2016:IPD:2839532.2840070} 
explore SQL offloading to smart, programmable storage devices. In this work, we focus on 
developing a more general approach towards compute offloading, which also includes SQL 
offloading. 

Summarizer offloads parts of computation on wimpy SSD ARM cores by enhancing the NVMe 
protocol~\cite{koo:micro17}. Willow~\cite{willow:osdi14} is a programmable SSD with an attached 
FPGA. Caribu\cite{istvan:vldb2017} implements ``smart distributed storage with replication'' in FPGAs, 
and provides a key-value interface with a support for iterations with predicates for 
compute offloading. IBM Netezza and Oracle Exadata also offload parts of the computation close to  
storage devices. In comparison, our approach focuses on synthesizing safe extensions that can be 
run in a shared storage environment. Naturally, any hardware accelerator can be used to accelerate 
the execution of the eBPF code (e.g., eBPF NIC offloading~\cite{ebpf-offload}). 
Splinter~\cite{splinter} is a recently proposed system that uses Rust language features to support the 
safe execution of user code in a multi-tenant KV store. Broadly speaking, sandboxes and fault isolation
research using compiler, language, or runtime systems is another promising 
avenue~\cite{sfi,Bershad:1995:ESP:224056.224077,Geambasu:2010:CAD:1924943.1924966,Gu:2016:BFN:3001136.3001154}. 
In this work, instead of starting from scratch, we focus on using the existing eBPF framework for 
developing and executing user-provided compute offloading extensions. 
Building advanced features in the storage devices like distributed shared log~\cite{corfu}, 
key-value stores~\cite{kv-ssd11,kv-ssd2}, higher-level abstractions and re-usable 
components~\cite{malacology,MacCormick:boxwood:2004,Coburn:2013:AMT:2517349.2522724} 
help to extend the utility of the storage service beyond just storing data. 
More recently, there is an open discussion on evaluating the suitability of eBPF for 
block devices on the Linux kernel mailing list as well~\cite{ebpf-blockdev}.
%

%
\section{Discussion Topics}

This paper discusses work which is still at early stages. We are fairly
confident that the ability to push operations to remote storage is important.
Hence, we would be interested in hearing objections to this premise, but also
potential applications that could benefit from this feature. We are less
confident that eBPF is best the way to tackle this issue, but we do think that
it will become increasingly used which means that building on it offers unique
benefits.  We would be interested in hearing counter-proposals to eBPF for
supporting these remote operations, as well as other ways of enhancing eBPF to
make it more suitable for data processing.

In our paper we focus on the challenge of safety and generality of using eBPF
for remote operations on disaggregated storage. A full system, however, needs
to address a number of other issues. We would be very interested in discussing
these issues, some of which we mention next.
NBD is a simple protocol and we used it as a placeholder, but a practical
solution will need to be piggybacked to more complicated protocols such as
NVMoF. Furthermore, another interesting question is how offloading to
disaggregated storage can be integrated with the rest of the storage stack,
and specifically how it can be used on top of a block device or even on top of
a file-system. Finally, allowing applications to upload arbitrary code has
non-obvious security implications (e.g., it may introduce side channels) and we
would need to understand the resulting security trade-offs.



\bibliographystyle{acm}
\bibliography{ms}

\end{document}